\documentclass[conference, 9pt]{IEEEtran}
\IEEEoverridecommandlockouts
\usepackage{cite}
\usepackage{amsmath,amssymb,amsfonts}
\usepackage{booktabs}
\usepackage{hyperref}
\usepackage{multirow}
\usepackage{algorithmic}
\usepackage{graphicx}
\usepackage{textcomp}
\usepackage{xcolor}
\usepackage{url}
\usepackage{colortbl}  
\usepackage{booktabs}
\usepackage{threeparttable}

\def\BibTeX{{\rm B\kern-.05em{\sc i\kern-.025em b}\kern-.08em
    T\kern-.1667em\lower.7ex\hbox{E}\kern-.125emX}}
\begin{document}

\title{\begin{minipage}[t]{\textwidth}
\centering
DRCap: Decoding CLAP Latents with Retrieval-Augmented Generation for Zero-shot Audio Captioning
\end{minipage}

\thanks{$^\dag$Qiuqiang Kong and Xie Chen are the corresponding authors.}
\thanks{$^\ast$Codes and models are available at \texttt{\scriptsize{\url{https://github.com/X-LANCE/SLAM-LLM/tree/main/examples/drcap_zeroshot_aac}. }}}

}

\author{
    \IEEEauthorblockN{
        Xiquan Li$^{1,2}$,
        Wenxi Chen$^1$, 
        Ziyang Ma$^1$, 
        Xuenan Xu$^1$, 
        Yuzhe Liang$^1$, 
        Zhisheng Zheng$^1$, \\
        Qiuqiang Kong$^{3\dag}$, 
        Xie Chen$^{1\dag}$
    }
    \IEEEauthorblockA{$^1$\textit{MoE Key Lab of Artificial Intelligence, X-LANCE Lab, Shanghai Jiao Tong University, China}}
    \IEEEauthorblockA{$^2$\textit{SJTU Paris Elite Institute of Technology, Shanghai Jiao Tong University, China}}
    \IEEEauthorblockA{$^3$\textit{Department of Electronics Engineering, The Chinese University of Hong Kong, China}}
}

\newcommand\blfootnote[1]{%
  \begingroup
  \renewcommand\thefootnote{}\footnote{#1}%
  \addtocounter{footnote}{-1}%
  \endgroup
}

\maketitle

\begin{abstract}
While automated audio captioning (AAC) has made notable progress, traditional fully supervised AAC models still face two critical challenges: the need for expensive audio-text pair data for training and performance degradation when transferring across domains. 
To overcome these limitations, we present DRCap, a data-efficient and flexible zero-shot audio captioning system that requires text-only data for training and can quickly adapt to new domains without additional fine-tuning. 
DRCap integrates a contrastive language-audio pre-training (CLAP) model and a large language model (LLM) as its backbone. 
During training, the model predicts the ground-truth caption with a fixed text encoder from CLAP, whereas, during inference, the text encoder is replaced with the audio encoder to generate captions for audio clips in a zero-shot manner. 
To mitigate the modality gap of the CLAP model, we use both the projection strategy from the encoder side and the retrieval-augmented generation strategy from the decoder side. 
Specifically, audio embeddings are first projected onto a text embedding support to absorb extensive semantic information within the joint multi-modal space of CLAP. 
At the same time, similar captions retrieved from a datastore are fed as prompts to instruct the LLM, incorporating external knowledge to take full advantage of its strong generative capability. 
Conditioned on both the projected CLAP embedding and the retrieved similar captions, the model is able to produce a more accurate and semantically rich textual description. 
By tailoring the text embedding support and the caption datastore to the target domain, DRCap acquires a robust ability to adapt to new domains in a training-free manner. 
Experimental results demonstrate that DRCap outperforms all other zero-shot models in in-domain scenarios and achieves state-of-the-art performance in cross-domain scenarios.
\end{abstract}


\section{Introduction}
Automated audio captioning (AAC) is a cross-modal translation task that seeks to generate natural language descriptions for given audio clips \cite{mei2022automated}. This process involves detailing the audio in terms of acoustic scenes, temporal relationships, object interactions, and environmental context \cite{xu2023beyond}.
Conventional AAC models often employ an encoder-decoder architecture \cite{sutskever2014sequence}, where an audio encoder extracts fine-grained audio features and a text decoder generates captions auto-regressively conditioned on these audio representations. The audio encoders used in previous studies \cite{kim2023prefix, pellegrini2023adapting, ghosh2024recap, wu2024improving, chen2024sjtu} are often pre-trained on tasks such as audio tagging or sound event detection \cite{kong2020panns, chen2022hts, chen2024eat},  while the text decoders are pre-trained large language models (LLMs) with extensive encyclopedic knowledge, such as BART \cite{lewis2019bart} or GPT-2 \cite{radford2019language}.

Despite the significant strides made in AAC, most fully supervised models still rely on extensively human-annotated datasets for training. However, data scarcity remains a critical issue for AAC, as annotating audio data demands careful attention and complex analysis. Moreover, given the diversity in audio concepts \cite{ghosh2024recap} and annotation styles across different datasets \cite{martin2021diversity}, existing fully supervised models often lack the flexibility to generalize to new domains, leading to diminished performance in cross-domain evaluations, where training and test data come from two distinct datasets. 

To overcome these challenges, researchers proposed zero-shot audio captioning framework \cite{deshmukh2024training, zhang2024zero, kouzelis2023weakly, salewski2023zero, shaharabany2023zero}, which seeks to generate captions without training on costly audio-text pair data. These works typically leverage the multi-modal capabilities of the CLAP model \cite{wu2023large, elizalde2023clap, elizalde2024natural}.
To bridge the modality gap \cite{liang2022mind} of CLAP, Deshmukh \textit{et al.} \cite{deshmukh2024training} and Kouzelis \textit{et al.} \cite{kouzelis2023weakly} injected Gaussian noise into CLAP latents, while Zhang \textit{et al.} \cite{zhang2024zero} crafted soft and hard prompts. However, adding noise can diminish the rich semantic information within the CLAP multi-modal space, while the fixed-category hard prompt in \cite{zhang2024zero} risks misleading the decoder.
Moreover, text decoders in previous works struggle to decode CLAP latents into accurate descriptions containing multiple sound events. 
A stronger LLM is required to fully leverage this joint multi-modal space.
As a result, although existing zero-shot audio captioning models demonstrate strong performance in cross-domain scenarios\cite{zhang2024zero}, they still lag significantly behind fully supervised models in in-domain scenarios.

\begin{figure*}[ht]
\vspace{-2mm}
\begin{minipage}[b]{1.0\linewidth}
  \centering
  \centerline{\includegraphics[width=17cm]{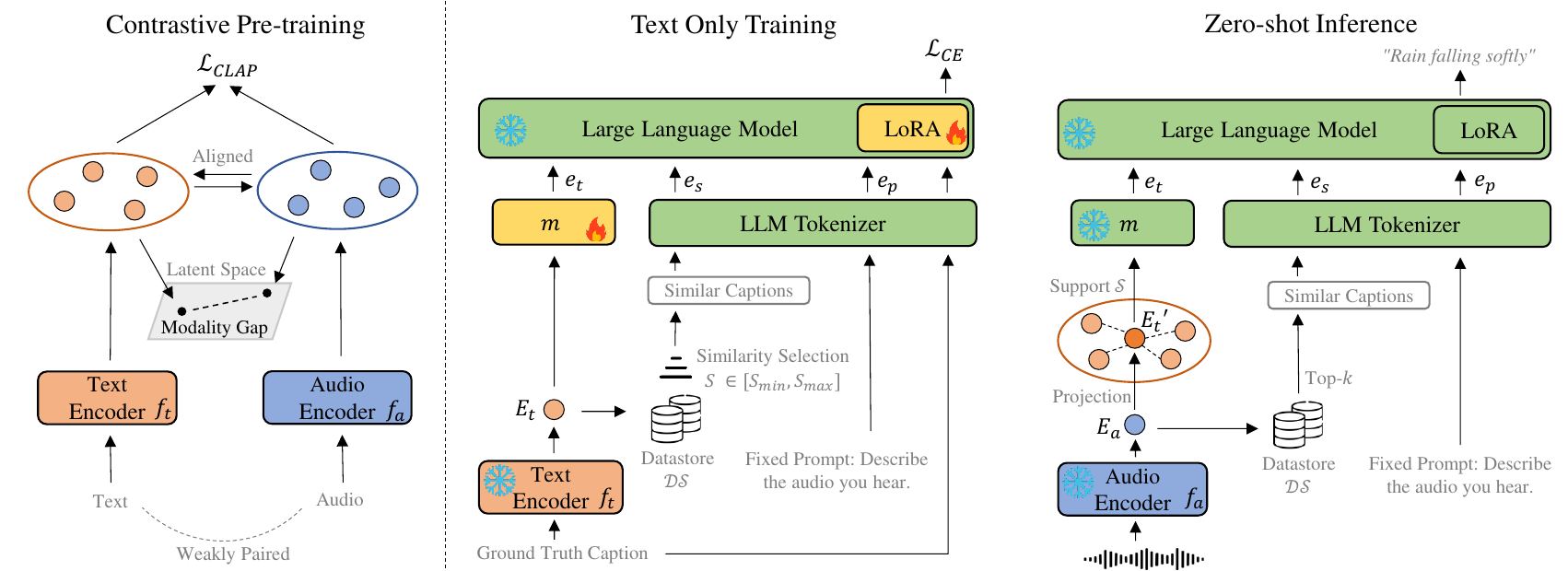}}
   \caption{
    Left: Overview of the CLAP model and the modality gap within its latent space. Right: Overview of the proposed DRCap. Based on the aligned multi-modal space of CLAP \cite{wu2023large}, during training, DRCap learns to decode the text embedding to reconstruct the original caption. Only the linear mapping network $m$ is trained, while the LLM is fine-tuned using the LoRA \cite{hu2021lora} method. The similarity selection was employed to prevent learning collapse caused by text-to-text retrieval. During inference, the audio embedding is first projected onto a text embedding support $\mathcal{S}$, mitigating the modality gap. The top-$k$ most similar captions retrieved from the datastore $\mathcal{DS}$ are used as prompts to instruct the LLM, producing accurate and semantically rich descriptions. Both the text embedding support $\mathcal{S}$ and the datastore $\mathcal{DS}$ could be changed in the inference stage, offering DRCap the flexibility to adapt to new domains.
       }
  \label{fig:DRCap model}
\end{minipage}
\vspace{-2mm}
\end{figure*}

In this paper, we propose DRCap, a data-efficient and transferable zero-shot audio captioning system that leverages the synergy between CLAP and LLM.
Based on the aligned multi-modal space of CLAP, DRCap requires only textual data for training, where a Vicuna-7B \cite{chiang2023vicuna} is fine-tuned with LoRA \cite{hu2021lora} to reconstruct the original caption from the CLAP text embedding. 
During inference, the text encoder is replaced by the audio encoder. 
To mitigate the modality gap and enhance the quality of the generated caption, we use both the projection strategy \cite{li2023decap} from the encoder side and the retrieval-augmented generation \cite{lewis2020retrieval} strategy from the decoder side. 
Specifically, during inference, audio embeddings are first projected onto a text embedding support, while semantically similar captions retrieved from an external datastore are fed as prompts to direct the LLM to create more accurate descriptions. 
Moreover, both the text embedding support and the caption datastore can be customized to match the target domain, providing our model with robust adaptability to new domains.
Experimental results demonstrate that DRCap performs comparably to fully supervised methods in in-domain scenarios and achieves state-of-the-art results in cross-domain scenarios.

\section{Methods}
\subsection{Overview}
We leverage the joint multi-modal space of CLAP to perform text-only training and then infer on audio clips in a zero-shot manner.
As illustrated in Figure \ref{fig:DRCap model} (left), CLAP jointly trains an audio encoder $f_a(\cdot)$ and a text encoder $f_t(\cdot)$ to align semantically similar audio-text pairs in a shared embedding space. After training, $f_a(a) \approx f_t(t)$ holds for any audio-text pair $(a, t)$. 

Given a raw caption $t \in \mathcal{T}$, where $\mathcal{T}$ represents a caption corpus, the objective in text-only training is to decode its CLAP text embedding $E_t = f_t(t)$ back into the original caption $t$. To achieve this, we train a lightweight linear mapping network $m$ to align the CLAP latent space with the LLM, producing $e_t=m(E_t)$. The LLM then reconstructs the original caption using $e_t$ along with an additional encoded prompt discussed in Section \ref{sec:rag}. During inference, given an audio clip $a$, the text encoder is replaced with the audio encoder of CLAP, extracting the audio embedding $E_a = f_a(a)$. 
Due to the modality gap between audio and text embeddings, directly feeding $E_a$ to the LLM through $m$ will yield sub-optimal results.
To address this issue and enhance the quality of generated captions, both the retrieval-augmented generation strategy and the projection-based decoding strategy are employed, detailed in Section \ref{sec:rag} and Section \ref{sec:projection} respectively. 

\subsection{Retrieval-Augmented Generation}
\label{sec:rag}
Retrieval-Augmented Generation (RAG) combines information retrieved from a datastore with a generative model, allowing it to generate more accurate, context-aware outputs by incorporating external knowledge.
DRCap leverages the RAG method to take full advantage of the generative abilities of the LLM, bridging the modality gap and improving its ability in describing unseen sound events. 

During training, given a raw caption $t$, we use its CLAP text embedding $E_t=f_t(t)$ to retrieve semantically similar captions from the datastore $\mathcal{DS}$. The retrieval process for a candidate caption $t_i \in \mathcal{DS}$ is based on the cosine similarity between their respective CLAP text embeddings, calculated as follows: 
\begin{equation}
S(t, t_\text{i}) = \frac{f_t(t) \cdot f_t(t_\text{i})}{\|f_t(t)\| \cdot \|f_t(t_\text{i})\|}
\end{equation}
We noticed, however, that naively selecting the top $k$ most similar captions can lead the LLM to become lazy, merely reproducing one of the $k$ captions as the output, neglecting $e_t$. To address this issue of learning collapse and improve the model's robustness, we proposed a similarity selection strategy, defining a similarity range $[S_\text{min}, S_\text{max}]$ from which $k$ captions are randomly selected. If fewer than $k$ captions fall within this range, only the qualifying captions are used as input. The effectiveness of our strategy is verified in Section \ref{sec:ablation}. 

Additionally, the model is given a fixed prompt (\textit{e.g.,} ``Describe the audio you hear") to help the LLM better understand the task. The similar captions and the fixed prompt are encoded using the tokenizer of the LLM. Let \(e_s \) and \( e_p \) denote the encoded embeddings of the similar captions and the fixed prompt.  The model is trained to minimize the cross-entropy loss conditioned on $z = \text{Concat}(e_t, e_s, e_p)$: 
\begin{equation}
\begin{aligned}
    \mathcal{L}_\text{CE} = -\frac{1}{L_t}\sum_{i=1}^{L_t}\log p(t_i|z, t_1, ..., t_{i-1})
\end{aligned}
\end{equation}
Where $e_t=m(f_t(t))$ is the mapped CLAP text embedding, $L_t$ is the length of the input caption $t$, $t_i$ is the $i$-th token of $t$. During training, we froze the CLAP encoder and trained only the linear mapping network, while applying LoRA \cite{hu2021lora} to fine-tune the large language model, which significantly enhanced training efficiency. 

During inference, given an audio clip $a$, the text-to-text retrieval is replaced with the audio-to-text retrieval, where we use the CLAP audio embedding $E_a=f_a(a)$ to retrieve $k$ most similar captions. The cross-modal similarity for a candidate caption $t_i \in \mathcal{DS}$ is defined as: 
\begin{equation}
    S(a, t_i) = \frac{f_a(a) \cdot f_t(t_i)}{\|f_a(a)\| \cdot \|f_t(t_i)\|} 
\end{equation}
The similarity selection is turned off, and instead, we choose the most similar captions to provide the LLM with maximum information.
\begin{table*}[t]
\centering
\caption{Performance comparison of AAC models for in-domain scenarios.}
\resizebox{1\linewidth}{!}{
\begin{tabular}{lllllllllllll}
\toprule
\multirow{3}{*}{\textbf{Method}} &   & \multicolumn{5}{c}{\textbf{Clotho} (\%)} & & \multicolumn{5}{c}{\textbf{AudioCaps} (\%)} \\
\cmidrule(lr){3-7} \cmidrule(lr){9-13}
 &  &  METEOR & CIDEr & SPICE & SPIDEr & FENSE && METEOR & CIDEr & SPICE & SPIDEr  & FENSE  \\
\hline
\rowcolor{gray!10}
\multicolumn{13}{c}{\textit{Fully Supervised Audio Captioning}}
\\
\hline
Prefix AAC \cite{kim2023prefix} & & 17.0 & 39.2 & 11.8 & 25.5 & - && 24.0 & 73.3 & 17.7 & 45.5  & -
\\
RECAP \cite{ghosh2024recap} &  & 17.7  & 41.1 & 12.5   & 22.4 & -  &  & \textbf{25.6} &  75.1  & 18.6 &  47.1  & -   \\
EnCLAP-large \cite{kim2024enclap} & & \textbf{18.2} & \textbf{42.6} & \textbf{12.9} & \textbf{27.8}  & \textbf{50.7}$^\dag$ && 25.5 & \textbf{80.3} & \textbf{18.8} & \textbf{49.5} & \textbf{65.5}$^\dag$ 
\\
\hline

\rowcolor{gray!10}
\multicolumn{13}{c}{\textit{Zero-shot Audio Captioning}}
\\
\hline
ZerAuCap \cite{salewski2023zero} && 9.4 & 14.0 & 5.3 & 9.7 & - && 12.3 & 28.1 & 8.6 & 18.3 & - \\
WSAC \cite{kouzelis2023weakly} && 17.4 & 37.1 & 12.3 & 24.7 & - && 24.1 & 63.3 & 17.3 & 40.3 & -\\
Zhang \textit{et al.} \cite{zhang2024zero} && 17.5 & 41.1 & 12.2 & 26.7 & 48.8 && 22.0 & 64.4 & 15.6 & 40.0 & - \\
DRCap (ours) &  & \textbf{18.2} & \textbf{43.8} & \textbf{13.3} & \textbf{28.5} & \textbf{53.0} && \textbf{25.3} & \textbf{70.5} & \textbf{18.0} & \textbf{44.2} &\textbf{66.2} \\
\textcolor{gray}{DRCap\textsubscript{LAION} (ours)$^\ddag$} & & \textcolor{gray}{17.9} & \textcolor{gray}{42.5} & \textcolor{gray}{12.4} & \textcolor{gray}{27.5} & \textcolor{gray}{51.9} & & \textcolor{gray}{25.7} & \textcolor{gray}{73.7} & \textcolor{gray}{17.9} & \textcolor{gray}{45.8} & \textcolor{gray}{66.4} \\
\bottomrule 
\end{tabular}
}
\begin{tablenotes}
\footnotesize 
\item $^\dag$: We evaluated metrics not reported in the original papers using the officially released checkpoint. $^\ddag$DRCap\textsubscript{LAION}: DRCap with LAION-CLAP as encoder. 
\end{tablenotes}
\label{tab:in-domain}
\vspace{-3mm}
\end{table*}

\subsection{Projection-based Decoding }
\label{sec:projection}
Moreover, during inference, instead of directly feeding the audio embedding $E_a$ to the LLM, we first project it onto the text embedding space of the CLAP model. 
Suppose that the system is trained on a caption corpus $\mathcal{T}=\{t_1, t_2, ..., t_N\}$, where $N$ denotes the size of $\mathcal{T}$. We can accumulate the text embeddings used during training, creating an embedding support $\mathcal{S}=\{E_{1}, E_{2}, ..., E_{N}\}$, where $E_{i}=f_t(t_i)$. For a given audio embedding $E_a$, its corresponding projected text-like embedding $E_t'$ could be obtained by performing a weighted combination of the text embeddings within the support:
 
\begin{equation}
    E_t' = \sum_{i=1}^{N}\frac{\exp\left((E_a^{\top} \cdot E_{i}) / \tau\right)}{\sum_{j=1}^{N} \exp\left((E_a^{\top} \cdot E_{j}) / \tau\right)} \cdot E_{i}
\end{equation}
Where $\tau$ is a temperature parameter, the projected vector $E_t'$ can capture the extensive semantic information from the support, while keeping its original acoustic features. $E_t'$ is then aligned with the LLM through the linear mapper $m$. Conditioned on both the projected CLAP embedding and the encoded similar captions, the LLM is able to generate more refined textual descriptions in a zero-shot manner.

\subsection{Domain Adaptation}
With the assistance of the text embedding support $\mathcal{S}$ and the caption datastore $\mathcal{DS}$, DRCap is capable of generating precise and meaningfully detailed captions. 
Moreover, the modifiability of both $\mathcal{S}$ and $\mathcal{DS}$ provides DRCap with the flexibility to quickly adapt to new domains. When encountering new sound event domains, relevant captions can be integrated into the text embedding support and the datastore. The multi-modal latent space of CLAP could then provide meaningful projected embeddings to decode, with similar captions guiding the LLM to describe the audio. Notably, no further training is needed for this entire process. The construction of $\mathcal{DS}$ and $\mathcal{S}$ will be discussed in Section \ref{sec:expsetup}. 

\section{Experimental Settings}

\subsection{Datasets}
\label{sec:datasets}
We train and evaluate DRCap on two most widely used AAC datasets, AudioCaps \cite{kim2019audiocaps} and Clotho \cite{drossos2020clotho}. AudioCaps is a subset of AudioSet \cite{gemmeke2017audio} that has been reannotated with caption labels. Each audio clip is annotated with a single caption for the training set and five captions for the validation and test sets. Our downloaded version contains 49274 examples for the training set, 494 for the validation set, and 957 for the test set.
Clotho consists of audio clips sourced from Freesound, each labeled with 5 captions. In our experiment, we use version 2.1 of Clotho, which contains 3839 examples in the training set, 1045 in the validation set, and 1045 in the test set.

The frozen CLAP model employed to extract audio and text embeddings is trained on WavCaps \cite{mei2024wavcaps} and Sound-VECaps \cite{yuan2024improving}. WavCaps comprises approximately 400k audio clips sourced from AudioSet-SL \cite{hershey2021benefit}, BBC Sound Effects,  FreeSound, and SoundBible, while Sound-VECaps contains approximately 1.6M audio clips sourced from AudioSet. Both datasets are weakly annotated with the assistance of ChatGPT \cite{schulman2022introducing}. We filtered out the audio clips in the dataset that overlap with AudioCaps or Clotho, assuming that the target-domain audio data are unavailable during training.

We also evaluated the performance of DRCap with the widely used CLAP model LAION-CLAP-630k\cite{wu2023large}, which is pre-trained on AudioCaps, Clotho and keyword-to-caption augmented AudioSet. Note that this scenario no longer qualifies as strict zero-shot audio captioning, as LAION-CLAP's pre-training has already utilized human-annotated audio-text pair data from AudioCaps and Clotho. This is also the reason why we opted to re-train a CLAP model using only weakly annotated data as described above.


\begin{table*}[t]
\centering
\caption{Performance comparison of AAC models for cross-domain scenarios. 
}
\vspace{1mm}
\resizebox{1\linewidth}{!}{
\begin{tabular}{lllllllllllll}
\toprule
\multirow{3}{*}{\textbf{Method}} &   & \multicolumn{5}{c}{\textbf{AudioCaps}  $\Longrightarrow$ \textbf{Clotho} (\%)} & & \multicolumn{5}{c}{\textbf{Clotho}  $\Longrightarrow$ \textbf{AudioCaps} (\%)} \\
\cmidrule(lr){3-7} \cmidrule(lr){9-13}
 &  &  METEOR & CIDEr & SPICE & SPIDEr & FENSE && METEOR & CIDEr & SPICE & SPIDEr  & FENSE  \\
\hline
\rowcolor{gray!10}
\multicolumn{13}{c}{\textit{Fully Supervised Audio Captioning}}
\\
\hline
EnCLAP-large$^\dag$ \cite{kim2024enclap} && 11.1 & 13.8 &5.9 & 9.9 & 36.1 && 13.3 & 17.4 & 8.0 & 12.6 & 38.8\\
Prefix AAC \cite{kim2023prefix}&  & 11.2 & 19.2 & 7.4 & 13.3 & - && 14.4 & 21.1 & 8.3 &14.7 & - 
 \\
RECAP \cite{ghosh2024recap} && \textbf{15.7} & \textbf{33.1} & \textbf{10.0} & \textbf{20.9} & -&& \textbf{16.9} & \textbf{35.7} & \textbf{11.1} & \textbf{20.4} & -\\
\hline
\rowcolor{gray!10}
\multicolumn{13}{c}{\textit{Zero-shot Audio Captioning}}
\\
\hline
WSAC$^\star$ \cite{kouzelis2023weakly} && 12.0 & 20.6 & 8.2 & 14.4 & - && 17.3 & 25.6 & 12.0 & 18.8 & -\\
Zhang \textit{et al.} \cite{zhang2024zero}&& 13.2 & 24.8 & 9.3 & 17.1 & - && 18.2 & 33.7 & 12.4 & 23.0 & 52.1\\
DRCap (ours) && \textbf{15.0} & \textbf{33.3} & \textbf{10.4} & \textbf{21.8} & \textbf{52.2} && \textbf{22.9} & \textbf{44.3} & \textbf{17.0} & \textbf{30.6}  & \textbf{62.6}  
\\
\textcolor{gray}{DRCap\textsubscript{LAION} (ours)}&  & \textcolor{gray}{17.3} & \textcolor{gray}{24.8} &\textcolor{gray}{12.3} &\textcolor{gray}{18.6} & \textcolor{gray}{48.2} & & \textcolor{gray}{21.7} &  \textcolor{gray}{45.4} & \textcolor{gray}{14.7} & \textcolor{gray}{30.0} & \textcolor{gray}{60.5} \\
\bottomrule
\end{tabular}
}
\begin{tablenotes}
\footnotesize 
\item  $^\dag$: Metrics evaluated on our test split using the officially released checkpoint.  $^\star$: Results provided by Zhang \textit{et al.}   \cite{zhang2024zero} based on their re-implementation. 
\end{tablenotes}
\label{tab:cross-domain}
\vspace{-3mm}
\end{table*}

\subsection{Experimental Setup}
\label{sec:expsetup}
To comprehensively evaluate the performance of DRCap, we conduct experiments in both in-domain and cross-domain setups: (1) We train and evaluate the model on the same dataset $\mathcal{D}_{source}$. (2) We train the model on the training set of $\mathcal{D}_{source}$, and evaluate on the test set of another dataset $\mathcal{D}_{target}$. During inference, for scenario (1), we use all text embeddings accumulated in the training stage as the support $\mathcal{S}$ mentioned in Section \ref{sec:projection}, which corresponds to the text embeddings of all captions in the training set of $\mathcal{D}_{source}$. For (2), we curate the text embedding support by encoding all captions from the training set of $\mathcal{D}_{target}$. In both settings, we use a caption datastore $\mathcal{DS}$ consisting of 450k captions sourced from WavCaps and the training sets of AudioCaps and Clotho. 


\subsection{Implementation Details}
DRCap was trained for 40,000 steps on AudioCaps and 20,000 steps on Clotho, with a peak learning rate of 1e-5, 1000 warm-up steps followed by a linear decay. We use the Adam optimizer \cite{kingma2014adam} and a batch size of 4. Validation was performed every 1000 steps, where the checkpoint with the lowest validation loss was saved for evaluation. Number of captions retrieved is set to $k=3$, and the range of similarity selection is fixed as $S_{min}=0.75, \ S_{max}=0.85$.
Our CLAP model, which employed the text encoder RoBERTa \cite{liu2019roberta} and the audio encoder HTS-AT \cite{chen2022hts}, was trained on WavCaps and Sound-VECaps, with a batch size of 256, a peak learning rate of 5e-5 for 15 epochs. Training followed a cosine annealing schedule with a 2-epoch warm-up phase, and the model of the last epoch was used. 


\section{Experimental Results}

\subsection{Main Results}

Tables \ref{tab:in-domain} and \ref{tab:cross-domain} present the performance of DRCap in both in-domain and cross-domain settings. We evaluated DRCap with our re-trained CLAP encoder or the widely used LAION-CLAP, denoted as DRCap and DRCap\textsubscript{LAION} respectively. Results of DRCap\textsubscript{LAION} are \textcolor{gray}{grayed out} for reference only, since it cannot be strictly considered as zero-shot audio captioning, as described in Section \ref{sec:datasets}. 
We compare DRCap’s performance with fully supervised AAC models: EnCLAP \cite{kim2024enclap}, Prefix-AAC \cite{kim2023prefix}, RECAP \cite{ghosh2024recap}, and zero-shot audio captioning models: ZerAuCap \cite{salewski2023zero}, WSAC \cite{kouzelis2023weakly} and Zhang \textit{et al.} \cite{zhang2024zero}. ZerAuCap \cite{salewski2023zero} uses CLAP to guide the LLM to generate descriptions, WSAC \cite{kouzelis2023weakly} trains a text decoder using the prefix language modeling paradigm conditioned on CLAP embeddings, while Zhang \textit{et al.} \cite{zhang2024zero} crafts soft and hard prompts to bridge the modality gap between audio and text embeddings of CLAP.  

Regardless of which CLAP model is employed, DRCap surpasses all competitive zero-shot audio captioning systems in in-domain scenarios by a large margin and is comparable with other fully-supervised methods. For cross-domain scenarios, it achieves state-of-the-art results across all metrics, highlighting its robust domain-transfer capability.
Furthermore, we found that DRCap outperforms other methods in terms of the FENSE \cite{zhou2022can} score in both two scenarios. We hypothesize that this advantage is due to DRCap’s ability to use the multi-modal space of CLAP, allowing it to  generate captions of better quality.

\subsection{Ablation Study}
\label{sec:ablation}
We conduct a comprehensive ablation study to validate each component of DRCap. 

\noindent\textbf{Similarity Selection} We turned off the similarity selection discussed in Section \ref{sec:rag}, instead selecting the top $k$ most similar captions during training.  
Since the ground-truth captions are available in the training stage, the retrieved most similar captions closely match the target in both semantics and vocabulary. This could lead the LLM to simply copy one of the retrieved captions as the output, trivializing the captioning task. However, during inference, without access to textual information, audio-to-text retrieval struggles to match the quality of text-to-text retrieval, and simply copying the retrieved captions hinders the model's performance, as shown in Table \ref{tab:ablation-in-domain} and Table \ref{tab:ablation-cross-domain}. Our proposed similarity selection strategy significantly alleviates the learning collapse and compels the LLM to take into account both the CLAP embedding and the retrieved captions, improving generation quality in both in-domain and cross-domain settings. 

\noindent\textbf{Retrieval-Augmented Generation. }We dropped all the similar captions in both the training and inference stages to evaluate the impact of RAG. As illustrated in table \ref{tab:ablation-in-domain} and table \ref{tab:ablation-cross-domain}, conditioning solely on CLAP embeddings results in inferior performance across all metrics in both scenarios, showing the advantage of similar captions in guiding the LLM to generate more accurate descriptions.

\noindent\textbf{LLM Fine-tuning. }We froze the LLM during training to conduct the ablation study on LoRA. Table \ref{tab:ablation-in-domain} and \ref{tab:ablation-cross-domain} highlight the significance of efficient LLM fine-tuning. Integrating LoRA adapters proved effective in aligning the CLAP latent space with the LLM.

\begin{table}[htbp]
\centering
\caption{Ablation Study of DRCap for in-domain scenarios}
\label{tab:ablation-in-domain}
\resizebox{1\linewidth}{!}{
\begin{tabular}{lccccc}
\toprule
\multirow{2.5}{*}{\textbf{Main Components}}  & \multicolumn{5}{c}{\textbf{AudioCaps} (\%)} \\ 
\cmidrule(lr){2-6}
& METEOR & CIDEr & SPICE & SPIDEr  & FENSE \\ 
\midrule
DRCap & \textbf{25.3} & \textbf{70.5} & 18.0 & \textbf{44.2} & \textbf{66.2}\\
\quad - w/o SS$^\dag$ & 21.8 & 59.5 & 15.7 & 37.6 & 61.8 \\
\quad - w/o RAG & 25.0 & 69.2 & \textbf{18.4} & 43.7 & 65.5 \\
\quad - w/o LoRA   & 23.7 & 64.7 & 16.4 & 40.5 & 64.1  \\ 
\quad - w/o PD$^\star$ & 19.9 & 31.2 & 13.4 & 22.3 & 55.4 \\ 
\bottomrule
\end{tabular}
}
\begin{tablenotes}
\footnotesize 
\item  $^\dag$: SS stands for similarity selection. $^\star$: PD for projection-based decoding.  
\end{tablenotes}
\vspace{-2mm}
\end{table}

\begin{table}[htbp]
\centering
\caption{Ablation Study of DRCap for cross-domain scenarios}
\label{tab:ablation-cross-domain}
\resizebox{1\linewidth}{!}{
\begin{tabular}{lccccc}
\toprule
\multirow{2.5}{*}{\textbf{Main Components}}  & \multicolumn{5}{c}{\textbf{AudioCaps $\Longrightarrow$ Clotho} (\%)} \\ 
\cmidrule(lr){2-6}
& METEOR & CIDEr & SPICE & SPIDEr  & FENSE \\ 
\midrule
DRCap & \textbf{15.0} & \textbf{33.3} & \textbf{10.4} & \textbf{21.8} & \textbf{52.2} \\
\quad - w/o SS$^\dag$ & 13.3 & 22.3 & 8.9& 15.6 & 46.0 \\
\quad - w/o RAG & 14.2 & 30.1 & 10.2 & 20.1 & 51.3 \\
\quad - w/o LoRA & 14.1 & 29.8 & 9.8 & 19.8 & 51.1    \\ 
\quad - w/o TD$^\star$ & 13.4 & 27.5 & 9.3 & 18.4 & 50.6 \\
\quad - w/o PD$^\ddag$  & 13.2 & 22.8 & 8.6 & 15.7 & 46.4 \\

\bottomrule
\end{tabular}
}
\begin{tablenotes}
\footnotesize 
\item  $^\dag$: SS denotes similarity selection. $^\star$: TD denotes target domain information. $^\ddag$: PD denotes projection-based decoding.
\end{tablenotes}
\vspace{-3mm}
\end{table}

\noindent\textbf{Projection-based Decoding}
We directly fed the audio embedding $e_a$ to the linear mapping network $m$ without using projection during inference to assess the benefit of projection-based decoding (PD). As illustrated in table \ref{tab:ablation-in-domain} and \ref{tab:ablation-cross-domain}, the modality gap caused a significant drop in performance when $e_a$ was used directly, while PD effectively bridge the discrepancy between audio and text embeddings. 

\noindent\textbf{Target Domain Information. }
We assume that no prior knowledge of the target domain is provided for cross-domain scenarios. Specifically, during inference, we use captions from the training set of $\mathcal{D}_{source}$ to construct $\mathcal{S}$, rather than using captions from the training set of $\mathcal{D}_{target}$. Furthermore, all captions from $\mathcal{D}_{target}$ are excluded from the datastore $\mathcal{DS}$. As shown in Table \ref{tab:ablation-cross-domain}, incorporating domain knowledge greatly improves DRCap’s cross-domain performance, demonstrating its adaptability during inference. 
Moreover, despite the absence of target domain knowledge, DRCap still performs competitively with SOTA methods, as shown in Table \ref{tab:cross-domain}.

\section{Conclusion and future work}
We present DRCap, a data-efficient and flexible audio captioning model that requires only textual data for training and can quickly adapt to other domains. 
Based on the CLAP model and the LLM, DRCap leverages projection-based decoding and retrieval-augmented generation to mitigate the modality gap. 
Conditioned on both the projected CLAP embedding and the retrieved similar captions, DRCap could produce more accurate and semantically rich descriptions. 
The replaceability of the text embedding support and the caption datastore guarantees the adaptability of the model. Experimental results show that DRCap outperforms other zero-shot audio captioning models in in-domain scenarios and achieves state-of-the-art performance in cross-domain scenarios. 


\clearpage

\section*{Acknowledgment}
This work was supported by the Science and Technology Innovation (STI) 2030-Major Project
(2022ZD0208700), the National Natural Science Foundation of China  (No. 62206171 and No. U23B2018), Shanghai Municipal Science and Technology Major Project under Grant 2021SHZDZX0102 and the International Cooperation Project of PCL.
\bibliographystyle{IEEEtran} 
\bibliography{refs}

\end{document}